\documentclass[aps,pre,showpacs]{revtex4}
\usepackage{epsf}
\usepackage{amsfonts}
\usepackage{eufrak}
\usepackage{amssymb}
\usepackage{graphicx}

\begin{document}

\title{Critical behaviour of ionic fluids}
\author{O.V. Patsahan}
\affiliation{Institute for Condensed Matter Physics of the
National Academy of
Sciences of Ukraine, \\
1 Svientsitskii Str., 79011 Lviv, Ukraine \\
E-mail: oksana@icmp.lviv.ua}
\date{\today}

\begin{abstract}
Recently we proposed a microscopic approach to the description of the
phase behaviour and critical phenomena in binary fluid mixtures. It was based on the
method of collective variables (CV) with a reference system.
The  approach allowed us to obtain the functional of the
Ginzburg-Landau-Wilson (GLW) Hamiltonian expressed in terms  of the
CV (fluctuating densities). The corresponding set of CV
included the variable connected with the order parameter. In this
paper we use this approach to the study of the  critical  behaviour of ionic fluids.
For the restricted primitive model (RPM) we obtain the functional of the grand partition
function in the phase space of the two fluctuating fields conjugate to the fluctuating densities.
First we calculate the phase diagram of the RPM  in the mean-field (MF) approximation
and then we do this calculation taking into account
the terms of the  higher orders in the effective Hamiltonian. In the both cases
the phase diagrams demonstrate  the gas-liquid (GL) and charge ordering
phase instabilities. In the latter case, the obtained value for the GL critical temperature
is in good agreement with the MC simulation data whereas the critical density is underestimated.
The explicit expression found for the grand thermodynamic potential in the vicinity of the
GL critical point implies a classical critical behaviour of the RPM.
\end{abstract}

\maketitle

\section{Introduction}

Nowdays the theory of phase transitions and critical phenomena is well
developed in general. It enables one to predict both universal and non-universal properties for
many model systems. However, a number of questions still remains open, among which
the critical behaviour of ionic fluids is of great interest.
Recent experiments have shown that both Ising-like and mean-field like
criticality can be observed in such systems. For reviews of the experimental
and theoretical situation see Refs. \cite{levelt1,pitzer,fisher1,fisher2,stell1,stell2}.
Numerous theoretical and computer simulation
studies of the restricted primitive model (RPM), the simplest model for ionic systems,
have not provided a clear picture of the thermodynamics in the critical region
\cite{zhou,fisher3,ciach1,ciachstell,caillol1,orkoulas1,caillol,luijten,panagiotopoulos1}.

In this paper  we address  the issue of the critical behaviour of the RPM using
the approach proposed in \cite{oksana,patkozmel} for the binary symmetrical mixture.
The theory has its origin in the approach based
on a functional representation of a partition function in the collective
variables (CV) space \cite{zubar,yukhol}. Its particular feature is a choice of
the phase space in which the system is considered. Among the independent variables
of this space there could be the ones connected with the order parameters. This
phase space is formed
by a set of CV. Each of them is a mode of density fluctuations corresponding to
the specificity of the model under consideration.  This approach
allows one to determine, on microscopic grounds, the explicit form of
the effective GLW Hamiltonian and then
to integrate the partition function in the neighborhood of the phase transition
point taking into account the renormalization group symmetry. As a result,
non-classical critical exponents and analytical expressions for thermodynamic
functions are obtained \cite{yuk,yuk2}.
More recently this theory was developed for
a binary fluid mixture \cite{oksana,patkozmel,patyuk3,patyuk4,patkozmel1,pat_pat}.

The paper consists of two parts. In the first part we obtain the
functional of the grand partition function (GPF) of the RPM given
in the phase space of the two fluctuating fields: the fluctuating
field $h_{{\bf{k}}=0}$ conjugate to the fluctuating total number
density $\rho_{{\bf{k}}=0}$ and the fluctuating field
$\gamma_{\bf{k}}$ conjugate to the fluctuating charge density
$c_{\bf{k}}$. Restricting our consideration to the second powers
of $\gamma_{\bf{k}}$ (taking into account the higher powers of
$h_{{\bf{k}}=0}$) we derive the equation for the chemical
potential. Based on the chemical potential obtained from the
linearized equation (that corresponds to the mean-field (MF)
approximation) we calculate the spinodal curve. Its run suggests
that two types of phase instabilities can occur in the RPM. One
part of the spinodal is   the gas-liquid (GL) type while another
one looks like a $\lambda$-line. We obtain the following values of
the GL critical point: $T_{c}^{*}=0.084$ and $\eta_{c}=0.005$
which agree with the other MF theories \cite{stell3}.
 In order to study the nature of the criticality of the RPM as well as to
get the best estimations for its GL critical point we go beyond
the above mentioned approximation taking into account the terms of
higher orders in the effective Hamiltonian. The second part of the
paper is devoted to this end. First, we consider the Gaussian
approximation of the functional of the GPF. It yields the equation
for the boundary of stability with respect to the charge density
fluctuations. Then, applying  the procedure proposed in
\cite{oksana,patkozmel}  we obtain the expression for the grand
thermodynamic potential in the vicinity of the GL critical point
as a power series in the field $\tilde h_{{\bf{k}}=0}$ (up to
$\tilde{h}^{4}$) conjugate to the order parameter. The expression
obtained has the form of the Landau free energy. We also calculate
the phase diagram demonstrating both GL and charge ordering phase
instabilities. The data for the GL critical point are
$T_{c}^{*}=0.0502$ and $\eta_{c}=0.022$.

\section{Functional representation of the GPF of the RPM. An equation for the chemical
potential; the phase disgram in the MF approximation}

The RPM consists of $N=N_{+}+N_{-}$ hard spheres of diameter $\sigma$
with $N_{+}$ carrying charges $+q$ and $N_{-}$ ($=N_{+}$) charges $-q$, in a medium
of dielectric constant $D$. The interaction potential of the RPM has the form
\begin{displaymath}
U_{\gamma\delta}(r) = \left\{\begin{array}{ll}
                     \infty &\mbox{if~~ $r<\sigma$}\\
                     \frac{q_{\gamma}q_{\delta}}{Dr}~~~~~&\mbox{if~~ $r\geq \sigma$}
                     \end{array}
              \right. , \quad  q_{i}=\pm q.
\end{displaymath}

We split the potential $U_{\gamma\delta}(r)$ into short- and long-range parts
using the Weeks-Chandler-Andersen partition \cite{wcha}. As a result,
we have
\begin{displaymath}
\psi_{\gamma\delta}(r) = \left\{\begin{array}{ll}
                     \infty &\mbox{if~~ $r\leq\sigma$}\\
                     0~~~~~&\mbox{if~~ $r> \sigma$}
                     \end{array}
              \right. ,
\end{displaymath}
%%%%%
\begin{displaymath}
\Phi_{\gamma\delta}(r) = \left\{\begin{array}{ll}
                     \frac{q_{\gamma}q_{\delta}}{D\sigma}&\mbox{if~~ $r\leq\sigma$}\\
                     \frac{q_{\gamma}q_{\delta}}{Dr}~~~~~&\mbox{if~~ $r> \sigma$}
                     \end{array}
              \right. .
\end{displaymath}

This simple form for $\Phi_{\gamma\delta}(r)$ inside the hard core changes the behaviour of the
Fourier transform  for large $k$ from usual Colombic $k^{-2}$ to $k^{-3}$ decay. As was shown
\cite{cha}, this choice of $\Phi_{\gamma\delta}(r)$ for $r<\sigma$ produces rapid convergence
of the series of the perturbation theory for the free energy.
The Fourier transform of $\Phi_{\gamma\delta}(r)=\frac{q^{2}}{Dr}=\Phi_{C}(r)$ has the form
\begin{equation}
\beta\rho\tilde \Phi_{C}(x)=24\beta^{*}\eta \frac{\sin x}{x^3},
\label{a2.1}
\end{equation}
where $\beta^{*}=\frac{\beta q^2}{D\sigma}$, $\beta=\frac{1}{k_{B}T}$,
$\eta=\frac{\pi}{6}\rho\sigma^3$ is fraction
density, $x=k\sigma$.

We start with the GPF. We use the formulas obtained within the framework of the CV method
for a binary continuous system (see (\ref{dA.4})-(\ref{dA.8}) in Appendix~A). In the case of
the RPM we have $\tilde V(k)\equiv 0$, $\tilde U(k)\equiv 0$, $\tilde W(k)=2\tilde \Phi_{C}(k)$
and $\mu_{1,+}=\mu_{1,-}$ ($\mu_{1,i}$ is a part of the chemical potential of the
$i$th species, $i=+,-$).
As a result, we can present the functional of the GPF of the RPM in the form
(for details see Appendix~A ):
\begin{equation}
\Xi=\Xi_{0}\int (d\rho)(dc)\exp\left(\beta\mu_{1}\rho_{0}-\frac{\beta}{V}\sum_{\bf{k}}
\tilde{\Phi}_{C}(k)c_{\bf{k}}c_{-\bf{k}}\right)J(\rho,c).
\label{a2.2}
\end{equation}
Here $\Xi_{0}$ is the GPF of the RS.
In the case of the RPM, the RS is
a one-component hard-sphere system with the diameter $\sigma$
(potential $\psi_{\gamma\delta}(r)=\psi(r)$).

$\rho_{\bf{k}}$ and $c_{\bf{k}}$ are the CV connected with total
density fluctuation modes and charge density fluctuation modes, respectively.
$\mu_{1}$ ($\mu_{1}=(\mu_{1,+}+\mu_{1,-})/\sqrt{2}$)
is determined from the equation
\begin{equation}
\frac{\partial\ln \Xi_{1}}{\partial\beta\mu_{1}} = \frac{\langle N\rangle}{\sqrt{2}}.
\label{a2.3}
\end{equation}

For the RPM $J(\rho,c)$ has the same form as that for the symmetrical binary fluid
\cite{patkozmel}:
\begin{eqnarray}
J(\rho,c)&=&\int (d\gamma)(d\omega)\exp\left\{\textit{i}2\pi\sum_{\bf{k}}(\omega_{\bf{k}}
\rho_{\bf{k}}+\gamma_{\bf{k}}c_{\bf{k}})
-\textit{i}2\pi\frac{1}{\sqrt{2}}\sum_{\bf{k}}{\mathfrak{M}}_{1}^{(0)}\omega_{\bf{k}}
\delta_{\bf{k}}
\right.
\nonumber \\
&&\left.+\frac{(-\textit{i}2\pi)^{2}}{2!}\frac{1}
{\sqrt{2}^{2}}\sum_{\bf{k}}({\mathfrak{M}}_{2}^{(0)}\omega_{\bf{k}}\omega_{-\bf{k}}
+{\mathfrak{M}}_{2}^{(2)}\gamma_{\bf{k}}\gamma_{-\bf{k}})+\frac{(-\textit{i}2\pi)^{3}}{3!}\frac{1}
{\sqrt{2}^{3}}
\right.
\nonumber \\
&&\left.\times\sum_{{\bf{k}}_{1},{\bf{k}}_{2},{\bf{k}}_{3}}({\mathfrak{M}}_{3}^{(0)}
\omega_{{\bf{k}}_{1}}\omega_{{\bf{k}}_{2}}\omega_{{\bf{k}}_{3}}
+3{\mathfrak{M}}_{3}^{(2)}\omega_{{\bf{k}}_{1}}\gamma_{{\bf{k}}_{2}}\gamma_{{\bf{k}}_{3}})
\delta_{{\bf{k}}_{1}+{\bf{k}}_{2}+{\bf{k}}_{3}}+\frac{(-\textit{i}2\pi)^{4}}{4!}
\right.
\nonumber \\
&&\left.\times\frac{1}
{\sqrt{2}^{4}}\sum_{{\bf{k}}_{1},\ldots,{\bf{k}}_{4}}
({\mathfrak{M}}_{4}^{(0)}\omega_{{\bf{k}}_{1}}\omega_{{\bf{k}}_{2}}\omega_{{\bf{k}}_{3}}
\omega_{{\bf{k}}_{4}}+6{\mathfrak{M}}_{4}^{(2)}\omega_{{\bf{k}}_{1}}\omega_{{\bf{k}}_{2}}
\gamma_{{\bf{k}}_{3}}\gamma_{{\bf{k}}_{4}}+{\mathfrak{M}}_{4}^{(4)}
\right.
\nonumber \\
&&\left.\times\gamma_{{\bf{k}}_{1}}\gamma_{{\bf{k}}_{2}}
\gamma_{{\bf{k}}_{3}}\gamma_{{\bf{k}}_{4}})\delta_{{\bf{k}}_{1}+\ldots+{\bf{k}}_{4}}
+\frac{(-\textit{i}2\pi)^{5}}{5!}\frac{1}
{\sqrt{2}^{5}}\sum_{{\bf{k}}_{1},\ldots,{\bf{k}}_{5}}
({\mathfrak{M}}_{5}^{(0)}\omega_{{\bf{k}}_{1}}\omega_{{\bf{k}}_{2}}\omega_{{\bf{k}}_{3}}
\right.
\nonumber \\
&&\left.\times
\omega_{{\bf{k}}_{4}}\omega_{{\bf{k}}_{5}}+10{\mathfrak{M}}_{5}^{(2)}\omega_{{\bf{k}}_{1}}
\omega_{{\bf{k}}_{2}}\omega_{{\bf{k}}_{3}}
\gamma_{{\bf{k}}_{4}}\gamma_{{\bf{k}}_{5}}+5
{\mathfrak{M}}_{5}^{(4)}\omega_{{\bf{k}}_{1}}
\gamma_{{\bf{k}}_{2}}\gamma_{{\bf{k}}_{3}}
\gamma_{{\bf{k}}_{4}}\gamma_{{\bf{k}}_{5}})
\right.
\nonumber \\
&&\left.\times\delta_{{\bf{k}}_{1}+\ldots+{\bf{k}}_{5}}+\frac{(-\textit{i}2\pi)^{6}}{6!}\frac{1}
{\sqrt{2}^{6}}\sum_{{\bf{k}}_{1},\ldots,{\bf{k}}_{6}}
({\mathfrak{M}}_{6}^{(0)}\omega_{{\bf{k}}_{1}}\omega_{{\bf{k}}_{2}}\omega_{{\bf{k}}_{3}}
\omega_{{\bf{k}}_{4}}\omega_{{\bf{k}}_{5}}\omega_{{\bf{k}}_{6}}
\right.
\nonumber \\
&&\left.+15{\mathfrak{M}}_{6}^{(2)}
\omega_{{\bf{k}}_{1}}\omega_{{\bf{k}}_{2}}\omega_{{\bf{k}}_{3}}\omega_{{\bf{k}}_{4}}
\gamma_{{\bf{k}}_{5}}\gamma_{{\bf{k}}_{6}}
+15{\mathfrak{M}}_{6}^{(4)}
\omega_{{\bf{k}}_{1}}\omega_{{\bf{k}}_{2}}\gamma_{{\bf{k}}_{3}}\gamma_{{\bf{k}}_{4}}
\gamma_{{\bf{k}}_{5}}\gamma_{{\bf{k}}_{6}})
\right.
\nonumber \\
&&\left.\times\delta_{{\bf{k}}_{1}+\ldots+{\bf{k}}_{6}}+\ldots
\right\}.
\label{a2.4}
\end{eqnarray}

In (\ref{a2.4}) the cumulants ${\mathfrak{M}}_{n}^{(i_{n})}$ with $i_{n}=0$ are connected
with the $n$th structure factors of the RS \cite{patyuk4}:
\[
{\mathfrak{M}}_{n}^{(0)}=\langle N\rangle S_{n}.
\]
Structure factors $S_{n}(0)$ with $n>2$ can be obtained from $S_{2}(0)$ by means of
a chain of equations for correlation functions \cite{stell}. Cumulants with $i_{n}\neq 0$
can be expressed in terms of ${\mathfrak{M}}_{n}^{(0)}$ (see also formulae (4.8) in
\cite{patyuk4}):
\begin{eqnarray}
{\mathfrak{M}}_{n}^{(2)}&=&{\mathfrak{M}}_{n-1}^{(0)}, \qquad
{\mathfrak{M}}_{n}^{(4)}=3{\mathfrak{M}}_{n-2}^{(0)}-2{\mathfrak{M}}_{n-3}^{(0)},
\nonumber \\
{\mathfrak{M}}_{n}^{(6)}&=&15{\mathfrak{M}}_{n-3}^{(0)}-30{\mathfrak{M}}_{n-4}^{(0)}+
16{\mathfrak{M}}_{n-5}^{(0)}.
\label{a2.4b}
\end{eqnarray}

First we integrate in (\ref{a2.2}) over the variables $\rho_{\bf{k}}$ and $\omega_{\bf{k}}$.
As a result, we have
\begin{equation}
\Xi=\Xi_{0}\exp\left(\sum_{n\geq 1}\frac{{\mathfrak{M}}_{n}^{(0)}}{n!}h_{0}^{n}\right)
\int (dc)\exp\left(-\frac{\beta}{V}\sum_{\bf{k}}
\tilde{\Phi}_{C}(k)c_{\bf{k}}c_{-\bf{k}}\right)J(c,h_{0}),
\label{a2.4a}
\end{equation}
%%%%%%%%%%%%%%%%%%%%%%%%%%%%%%%%%%%%%%%%%
where
\begin{eqnarray}
J(c,h_{0})&=& \int (d\gamma)\exp\left(\textit{i}2\pi\sum_{\bf{k}}\gamma_{\bf{k}}c_{\bf{k}}
+\frac{(-\textit{i}2\pi)^{2}}{2!}\frac{1}
{\sqrt{2}^{2}}\sum_{\bf{k}}\gamma_{\bf{k}}\gamma_{-\bf{k}}\sum_{n\geq 0}{\mathfrak{M}}_{n+2}^{(2)}
\frac{h_{0}^{n}}{n!}
\right.
\nonumber \\
&&\left.
+\frac{(-\textit{i}2\pi)^{4}}{4!}\frac{1}
{\sqrt{2}^{4}}\sum_{{\bf{k}}_{1},\ldots,{\bf{k}}_{4}}
\gamma_{{\bf{k}}_{1}}\gamma_{{\bf{k}}_{2}}
\gamma_{{\bf{k}}_{3}}\gamma_{{\bf{k}}_{4}}\delta_{{\bf{k}}_{1}+\ldots+{\bf{k}}_{4}}
\sum_{n\geq 0}{\mathfrak{M}}_{n+4}^{(4)}\frac{h_{0}^{n}}{n!}\right),
\label{a2.5}
\end{eqnarray}
%%%%%%%%%%%%%%%%%%%%%%%%%%%%%%%%%%%%%%55
\begin{equation}
h_{\bf{k}=0}=h_{0}=\frac{\beta\mu_{1}}{\sqrt{2}}
\label{a2.6}
\end{equation}
and condition (\ref{a2.3}) has the form:
\begin{equation}
\frac{\partial\ln \Xi_{1}}{\partial h_{0}} = \langle N\rangle.
\label{a2.7}
\end{equation}
Expressions  (\ref{a2.4a})-(\ref{a2.5}) do not include the "field"
variable $h_{\bf{k}}$ with $\bf{k}\neq 0$. As one can see below,
this fact will give rise to the Landau type free energy of the RPM
in the vicinity of the GL critical point.

Performing in (\ref{a2.4a})-(\ref{a2.5}) integration over $c_{\bf{k}}$ we obtain
(up to $\gamma^{4}$)
\begin{eqnarray}
\Xi_{1}&=&\prod_{\bf{k}}\sqrt{\frac{\pi V}{\beta|\tilde{\Phi}_{C}(k)|}}\int (d\gamma)
\exp\left\{{\mathcal{F}}(h_{0})+\frac{(-\textit{i}2\pi)^{2}}{2!}
\sum_{\bf{k}}\gamma_{\bf{k}}\gamma_{-\bf{k}}\frac{1}{2\frac{\beta}{V}\tilde{\Phi}_{C}(k)}
\right.
\nonumber\\
&&\left.
\times\left(1+\frac{\beta}{V}\tilde{\Phi}_{C}(k)\frac{\partial {\mathcal{F}}(h_{0})}
{\partial h_{0}}\right)
+\frac{(-\textit{i}2\pi)^{4}}{4!}\frac{1}
{\sqrt{2}^{4}}\sum_{{\bf{k}}_{1},\ldots,{\bf{k}}_{4}}
\gamma_{{\bf{k}}_{1}}\gamma_{{\bf{k}}_{2}}
\gamma_{{\bf{k}}_{3}}\gamma_{{\bf{k}}_{4}}
\right.
\nonumber \\
&&\left.
\times
\delta_{{\bf{k}}_{1}+\ldots+{\bf{k}}_{4}}
\left(3\frac{\partial^{2} {\mathcal{F}}(h_{0})}{\partial h_{0}^{2}}-2
\frac{\partial {\mathcal{F}}(h_{0})}{\partial h_{0}}\right)\right\},
\label{a2.8}
\end{eqnarray}
where
\begin{equation}
{\mathcal{F}}(h_{0})=\sum_{n\geq 1}\frac{{\mathfrak{M}}_{n}^{(0)}}{n!}h_{0}^{n}, \qquad
\frac{\partial {\mathcal{F}}(h_{0})}{\partial h_{0}}=\sum_{n\geq 1}
\frac{{\mathfrak{M}}_{n}^{(0)}}{(n-1)!}h_{0}^{n-1},
\label{a2.10a}
\end{equation}
\begin{equation}
\frac{\partial^{2} {\mathcal{F}}(h_{0})}{\partial h_{0}^{2}}=\sum_{n\geq 2}
\frac{{\mathfrak{M}}_{n}^{(0)}}{(n-2)!}h_{0}^{n-2}
\label{a2.11a}
\end{equation}
and formulas (\ref{a2.4b})  are used for ${\mathfrak{M}}_{n}^{(i_{n})}$.

It should be point out that this integration makes sense for a
positive function $\tilde \Phi_{C}(k)$. In our study $\tilde
\Phi_{C}(k)$ (see (\ref{a2.1})) can change the sign and the
transition from (\ref{a2.4a})-(\ref{a2.5}) to (\ref{a2.8}) has to
be regarded as an algebraic formal way of deriving the expansion
in terms of $h_{0}$ and  $\gamma_{\bf{k}}$ solely. As a result, in
(\ref{a2.8}) in the exponent we have the expression in terms of
the two fluctuating fields,namely, the fluctuating field
$h_{{\bf{k}}=0}$ conjugate to the fluctuating total density
$\rho_{{\bf{k}}=0}$ and the fluctuating field $\gamma_{\bf{k}}$
(for all of $\bf{k}$) conjugate to the fluctuating charge density
$c_{\bf{k}}$.

\paragraph{Equation for the chemical potential.}
Restricting our consideration in (\ref{a2.5}) to the second power of $\gamma_{\bf{k}}$ and
assuming $\frac{\partial {\mathcal{F}}(h_{0})}{\partial h_{0}}>0$
(${\mathfrak{M}}_{n+2}^{(2)}\frac{h_{0}^{n}}{n!}>0$), we integrate in
(\ref{a2.4a})-(\ref{a2.5}) over $\gamma_{\bf{k}}$ and $c_{\bf{k}}$. As a result, we
obtain for  $\Xi_{1}$
\begin{equation}
\Xi_{1}=\prod_{\bf{k}}\left[1+\frac{\beta}{V}\tilde{\Phi}_{C}(k)
\frac{\partial {\mathcal{F}}(h_{0})}{\partial h_{0}}
\right]^{-1/2}\exp\left({\mathcal{F}}(h_{0})\right).
\label{a2.10}
\end{equation}

Using (\ref{a2.7}) and (\ref{a2.10}) we can write the following equation for the chemical
potential
\begin{equation}
\frac{1}{2}\sum_{\bf{k}}\frac{\frac{\beta}{V}\tilde{\Phi}_{C}(k)}{1+\frac{\beta}{V}
\tilde{\Phi}_{C}(k)\frac{\partial {\mathcal{F}}(h_{0})}{\partial h_{0}}}=
\frac{\frac{\partial {\mathcal{F}}(h_{0})}{\partial h_{0}}-{\mathfrak{M}}_{1}^{(0)}}
{\frac{\partial^{2}
{\mathcal{F}}(h_{0})}{\partial h_{0}^{2}}}=
h_{0}\left(1-\frac{1}{2}
\frac{{\mathfrak{M}}_{3}^{(0)}}
{{\mathfrak{M}}_{2}^{(0)}}h_{0}+\ldots\right),
\label{a2.11}
\end{equation}
where $h_{0}$, ${\mathcal{F}}(h_{0})$, $\frac{\partial {\mathcal{F}}(h_{0})}{\partial h_{0}}$
and $\frac{\partial^{2} {\mathcal{F}}(h_{0})}{\partial h_{0}^{2}}$ are given by (\ref{a2.6})
and (\ref{a2.10a})-(\ref{a2.11a}), respectively.

\paragraph{Phase diagram  of the RPM in the MF approximation.}
We solve equation (\ref{a2.11}) in the simplest approximation. Neglecting terms
$h_{0}^{2}$, $h_{0}^{3}$ and etc. in the right hand side of (\ref{a2.11}) and setting
\begin{equation}
\frac{\partial {\mathcal{F}}(h_{0})}{\partial h_{0}}={\mathfrak{M}}_{1}^{(0)}
\label{a3.1a}
\end{equation}
in the left hand
side of (\ref{a2.11}) we obtain for $\mu_{1,+}$
($=\mu_{1,-}$)
\begin{equation}
\mu_{1,+}=-\frac{1}{2\beta}\tilde{a}(\beta),
\label{a3.1}
\end{equation}
where
\begin{equation}
\tilde{a}(\beta)=-\sum_{\bf{k}}\frac{\frac{\beta}{V}\tilde{\Phi}_{C}(k)}{1+\frac{\beta}{V}
\tilde{\Phi}_{C}(k){\mathfrak{M}}_{1}^{(0)}}.
\label{a3.2}
\end{equation}

On the other hand, we can get the same result for $\mu_{1,+}$
taking into account  the terms proportional to $\gamma^{2}$ and
$\gamma^{2}h_{0}$ in the exponent of (\ref{a2.5}) and  setting
${\mathfrak{M}}_{3}^{(2)}\equiv 0$ in the final result
(${\mathfrak{M}}_{3}^{(2)}$ is the coefficient of
$\gamma^{2}h_{0}$). Therefore, we may say that $\mu_{1,+}$ given
by (\ref{a3.1})-(\ref{a3.2}) is obtained in the MF approximation.

The full chemical potential $\mu_{+}$ is equal to \cite{patyuk3}:
\[
\mu_{+}=\mu_{0,+}+\mu_{1,+},
\]
where $\mu_{0,+}$ ($=\mu_{0,-}$) is the chemical potential of a one-component
hard sphere system.

The equation $\beta\rho\frac{\partial\mu_{+}}{\partial\rho}=0$, where $\rho$
is the total number density, gives the spinodal of the RPM in the approximation
considered. Using (\ref{a2.1}) for $\tilde{\Phi}_{C}(k)$ and the Perkus-Yevick
approximation for the RS, the equation for the spinodal curve can be written as
\begin{equation}
\int_{0}^{\infty}\frac{x^{2}(\sin x)^{2}dx}{(x^{3}T^{*}+24\eta\sin x)^{2}}
=\frac{\pi}{24\eta}\frac{(1+2\eta)^{2}}{(1-\eta)^{4}}
\label{a3.3}
\end{equation}

The phase diagram calculated from (\ref{a3.3}) is shown in Fig.~1. The obtained data
for the GL critical point are $T_{c}^{*}\simeq 0.084$ ($T_{c}^{*}=1/\beta_{c}^{*}$)
and $\eta_{c}\simeq 0.005$.

It is evident that in the considered approximation (corresponding to the MF theory) we
get the overestimated value for
the GL critical temperature and the underestimated value for the critical density.
But, in contrast to the previous results, the spinodal curve
changes its run (at $\eta\simeq 0.047$) and then it directs to the higher temperature.
The second positive slop of the spinodal indicates another type of the phase instability
appearing in the RPM (similarly to the $\lambda$-line in a symmetrical binary fluid).
Below, we shall calculate the phase diagram of the RPM taking into account in (\ref{a2.5})
the terms of the higher orders.

\section{Criticality of the RPM: beyond the MF theory}
\setcounter{paragraph}{0}
Now let us rewrite (\ref{a2.4a})-(\ref{a2.5}) as
\begin{eqnarray}
\Xi_{1}&=&\exp\left({\mathcal{F}}(h_{0})\right)
\int (dc)\exp\left(-\frac{\beta}{V}\sum_{\bf{k}}
\tilde{\Phi}_{C}(k)c_{\bf{k}}c_{-\bf{k}}\right)
\nonumber \\
&&\times\int (d\gamma)
\exp\left\{\textit{i}2\pi\sum_{\bf{k}}\gamma_{\bf{k}}c_{\bf{k}}
+\frac{(-\textit{i}2\pi)^{2}}{2!}\frac{1}
{\sqrt{2}^{2}}\sum_{\bf{k}}\gamma_{\bf{k}}\gamma_{-\bf{k}}{\mathfrak{M}}_{1}^{(0)}
\right.
\nonumber \\
&&\left.
+\frac{(-\textit{i}2\pi)^{2}}{2!}\frac{1}
{\sqrt{2}^{2}}\sum_{\bf{k}}\gamma_{\bf{k}}\gamma_{-\bf{k}}\left(\frac{\partial {\mathcal{F}}(h_{0})}
{\partial h_{0}}-{\mathfrak{M}}_{1}^{(0)}\right)
+\frac{(-\textit{i}2\pi)^{4}}{4!}\frac{1}{\sqrt{2}^{4}}
\right.
\nonumber \\
&&\left.
\times\sum_{{\bf{k}}_{1},\ldots,{\bf{k}}_{4}}
\gamma_{{\bf{k}}_{1}}\gamma_{{\bf{k}}_{2}}
\gamma_{{\bf{k}}_{3}}\gamma_{{\bf{k}}_{4}}
\left(3\frac{\partial^{2} {\mathcal{F}}(h_{0})}{\partial h_{0}^{2}}-
2\frac{\partial {\mathcal{F}}(h_{0})}
{\partial h_{0}}\right)\delta_{{\bf{k}}_{1}+\ldots+{\bf{k}}_{4}}\right\}
\label{a3.4}
\end{eqnarray}
and  the notations in (\ref{a3.4}) are the same as those in the previous section.

\paragraph{Gaussian approximation.}
 First, we restrict our attention to the second powers
of variables $h_{0}$ and $\gamma_{\bf{k}}$ which corresponds to
neglecting the terms proportional to $h_{0}^{3}$,
$\gamma_{\bf{k}}\gamma_{-\bf{k}}h_{0}$,
$\gamma_{\bf{k}}\gamma_{-\bf{k}}h_{0}^{2}$, $h_{0}^{4}$, etc.
After integration in (\ref{a3.4}) over $\gamma_{\bf{k}}$ we obtain
\begin{eqnarray}
\Xi_{1}&=&\exp\left({\mathcal{F}}(h_{0})\right)\prod_{\bf{k}}\left(\pi{\mathfrak{M}}_{1}^{(0)}
\right)^{-1/2}
\int (dc)\exp\left\{-\sum_{\bf{k}}\frac{c_{\bf{k}}c_{-\bf{k}}}{{\mathfrak{M}}_{1}^{(0)}}
\right.
\nonumber \\
&&\left.
\times\left(1+\frac{\beta}{V}
\tilde{\Phi}_{C}(k){\mathfrak{M}}_{1}^{(0)}\right)\right\}.
\label{a3.5}
\end{eqnarray}
As is seen from (\ref{a3.5}), the equality
\begin{equation}
1+\frac{\beta}{V}
\tilde{\Phi}_{C}(k){\mathfrak{M}}_{1}^{(0)}=0
\label{a3.6}
\end{equation}
holds at some values of the wave-vector $\bf{k}$, temperature and density.
Equation (\ref{a3.6}) determines the boundary of stability connected with the charge
fluctuations (the field variable
$\gamma_{\bf{k}}$ is conjugate to the CV $c_{\bf{k}}$):
\begin{equation}
T^{*}=-24\eta \frac{\sin x}{x^3}, \qquad T^{*}=\frac{1}{\beta^{*}},
\label{a3.6a}
\end{equation}
or
\begin{equation}
T_{c}^{*}(x^{*},\eta)=-8\eta \frac{\\cos x^{*}}{{x^{*}}^2},
\label{a3.6b}
\end{equation}
where $x^{*}$ is determined from the condition
\begin{equation}
\tan x^{*}=\frac{x^{*}}{3},
\label{a3.6c}
\end{equation}
which yields $x^{*}\simeq 4.0783$. Substituting $x^{*}$ in (\ref{a3.5}) we obtain the
boundary of stability with respect to fluctuations of the local charge density
\begin{equation}
T_{c}^{*}(x=x^{*})\simeq 0.285\eta.
\label{a3.7}
\end{equation}
A similar result (for another choice of interaction inside the
hard core) was obtained in \cite{ciach1,ciachstell} within the
framework of the field-theoretical approach. The possibility of
the charge-ordering transition in the continuous-space RPM model
was discussed therein.

It is worth noting that the RPM does not demonstrate the GL phase
transition in the Gaussian approximation. In order to obtain the
GL spinodal curve we should take into consideration the terms of
the order higher than the second one ($\gamma^{2}h$,
$\gamma^{2}h^{2}$, etc.).

\paragraph{Beyond the Gaussian approximation(model $\varphi^{4}$).}

We restrict our consideration in (\ref{a3.4}) to the terms of the fourth order.
In this case $\Xi$ has the form:
\begin{eqnarray}
\Xi&=&\Xi_{0}\exp\left({\mathcal{F}}(h_{0})\right)
\int (dc)\exp\left(-\frac{\beta}{V}\sum_{\bf{k}}\tilde{\Phi}_{C}(k)c_{\bf{k}}c_{-\bf{k}}
\right)\int (d\gamma)\exp\left\{\textit{i}2\pi
\right.
\nonumber \\
&&\left.
\times
\sum_{\bf{k}}\gamma_{\bf{k}}c_{\bf{k}}
+\frac{(-\textit{i}2\pi)^{2}}{2!\sqrt{2}^{2}}
 \sum_{\bf{k}}\gamma_{\bf{k}}\gamma_{-\bf{k}}\left({\mathfrak{M}}_{1}^{(0)}
+h_{0}{\mathfrak{M}}_{2}^{(0)}+h_{0}^{2}{\mathfrak{M}}_{3}^{(0)}\right)
\right.
\nonumber \\
&&\left.
+\frac{(-\textit{i}2\pi)^{4}}{4!\sqrt{2}^{4}}
\sum_{{\bf{k}}_{1},\ldots,{\bf{k}}_{4}}
\gamma_{{\bf{k}}_{1}}\gamma_{{\bf{k}}_{2}}
\gamma_{{\bf{k}}_{3}}\gamma_{{\bf{k}}_{4}}
\left(3{\mathfrak{M}}_{2}^{(0)}-2{\mathfrak{M}}_{1}^{(0)}\right)
\delta_{{\bf{k}}_{1}+\ldots+{\bf{k}}_{4}}\right\},
\label{a3.8}
\end{eqnarray}
where the summation in (\ref{a2.10a}) over $n$ is restricted to 4.

Now we follow the programme proposed in \cite{patyuk4,oksana} for a two-component
fluid system. First, we separate the two types of variables: the essential variables
(which include the variable connected with the order parameter) and the
non-essential variables. Then, integrating over the non-essential variables with the
Gaussian density measure, we construct the
basic density measure (the GLW Hamiltonian) with respect to the essential variables.

For the RPM in the vicinity of the GL critical point the variable
$h_{0}$ (conjugate to the CV $\rho_{0}$) turns out to be the
essential variable \cite{patkozmel}. Thus, we can present
(\ref{a3.8}) as
\begin{eqnarray}
\Xi_{1}&=&\prod_{\bf{k}}\left(\pi{\mathfrak{M}}_{1}^{(0)}
\right)^{-1/2}\exp\left({\mathcal{F}}(h_{0})\right)
\int (dc)\exp\left(-\frac{\beta}{V}\sum_{\bf{k}}
\tilde{\Phi}_{C}(k)c_{\bf{k}}c_{-\bf{k}}\right)
\nonumber \\
&&\times
\left(1+\hat{\mathcal{A}}+\frac{1}{2!}\hat{\mathcal{A}}^{2}+\ldots\right)
\exp\left(-\sum_{\bf{k}}c_{\bf{k}}c_{-\bf{k}}/{\mathfrak{M}}_{1}^{(0)}\right),
\label{a3.9}
\end{eqnarray}
where
\[
\hat{\mathcal{A}}=\left(h_{0}{\mathfrak{M}}_{2}^{(0)}+h_{0}^{2}{\mathfrak{M}}_{3}^{(0)}\right)
\frac{\partial ^{2}}{\partial c_{\bf{k}}c_{-\bf{k}}}.
\]
After integration in (\ref{a3.9}) we get
\begin{equation}
\Xi_{1}=\prod_{\bf{k}}\frac{1}{\sqrt{1+\beta\frac{\langle N\rangle}{V}
\tilde{\Phi}_{C}(k)}}
\exp\left(\sum_{n\geq 1}{{\mathcal{M}}}_{n}\frac{h_{0}^{n}}{n!}\right),
\label{a3.10}
\end{equation}
where
\[
{{\mathcal{M}}}_{n}={\mathfrak{M}}_{n}^{(0)}+\Delta {\mathfrak{M}}_{n},
\]
$\Delta {\mathfrak{M}}_{n}$ are the corrections obtained as the result of integration over
variables $c_{\bf{k}}$:
\begin{eqnarray}
\Delta{\mathfrak{M}}_{1}&=&\frac{1}{2}{\mathfrak{M}}_{2}^{(0)}\tilde{a}(\beta),
\quad
\Delta{\mathfrak{M}}_{2}=\frac{1}{2}{\mathfrak{M}}_{3}^{(0)}\tilde{a}(\beta),
\nonumber \\
\Delta{\mathfrak{M}}_{3}&\equiv& 0, \qquad \Delta{\mathfrak{M}}_{4}\equiv 0.
\label{a3.11}
\end{eqnarray}
and $\tilde{a}(\beta)$ is given by (\ref{a3.2}).

It is worth noting that the non-zero corrections $\Delta
{\mathfrak{M}}_{n}$ (see (\ref{a3.11})) are those which include
only one sum over $\bf{k}$.

Next, the shift is carried out in order to eliminate the cubic term in (\ref{a3.10})
\[
h_{0}=\tilde{h_{0}}+\Delta,
\]
where $\Delta=-\frac{{\mathcal{M}}_{3}}{{\mathcal{M}}_{4}}$. Then (\ref{a3.10}) has the form
\begin{equation}
\Xi_{1}={\mathcal{C}}\exp\left(\sum_{n\geq 1}\tilde{{\mathcal{M}}}_{n}\frac{\tilde{h_{0}}^{n}}
{n!}\right),
\label{a3.12}
\end{equation}
where
\begin{eqnarray*}
{\mathcal{C}}&=&\prod_{\bf{k}}\frac{1}{\sqrt{1+\beta\frac{\langle N\rangle}{V}
\tilde{\Phi}_{C}(k)}}\exp\left({\mathfrak{M}}_{1}^{(0)}+\frac{1}{2}
\left(\frac{{\mathfrak{M}}_{3}^{(0)}}{{\mathfrak{M}}_{4}^{(0)}}\right)^{2}
\left({\mathfrak{M}}_{2}^{(0)}-\frac{({\mathfrak{M}}_{3}^{(0)})^{2}}{4{\mathfrak{M}}_{4}^{(0)}}\right)
\right.
\\
&&\left.
+\frac{1}{2}\tilde{a}(\beta)
\left({\mathfrak{M}}_{2}^{(0)}+\frac{({\mathfrak{M}}_{3}^{(0)})^{3}}{2({\mathfrak{M}}_{4}^{(0)})^{2}}
\right)
\right),
\\
\tilde{{\mathcal{M}}}_{1}&=&
{\mathfrak{M}}_{1}^{(0)}-\frac{{\mathfrak{M}}_{3}^{(0)}}{{\mathfrak{M}}_{4}^{(0)}}
\left({\mathfrak{M}}_{2}^{(0)}-\frac{({\mathfrak{M}}_{3}^{(0)})^{2}}{3{\mathfrak{M}}_{4}^{(0)}}\right)
%\\
%&&
+\frac{1}{2}\tilde{a}(\beta)
\left({\mathfrak{M}}_{2}^{(0)}-\frac{({\mathfrak{M}}_{3}^{(0)})^{2}}{{\mathfrak{M}}_{4}^{(0)}}
\right),
\\
\tilde{{\mathcal{M}}}_{2}&=&{\mathfrak{M}}_{2}^{(0)}-\frac{({\mathfrak{M}}_{3}^{(0)})^{2}}
{2{\mathfrak{M}}_{4}^{(0)}}+\frac{1}{2}\tilde{a}(\beta){\mathfrak{M}}_{3}^{(0)}, \qquad
\tilde{{\mathcal{M}}}_{4}={\mathfrak{M}}_{4}^{(0)}.
\end{eqnarray*}
%%%%%%%
In (\ref{a3.12}) $\tilde{h_{0}}$ is the field conjugate to the order parameter
for the GL critical point (see (1.8)-(1.10)
in \cite{yuk2} and (15)-(20) in \cite{patkozmel}).

Now we can obtain the grand thermodynamic potential of the RPM in the vicinity of the
GL critical point
\begin{equation}
\Omega-\Omega_{0}+k_{B}T\ln {\mathcal{C}} = -k_{B}T\left(\sum_{n\geq 1}
\tilde{{\mathcal{M}}}_{n}\frac{\tilde{h_{0}}^{n}}{n!}\right).
\label{a3.13}
\end{equation}
The right hand side in (\ref{a3.13}) has the form of the Landau free energy
expressed in terms of the field $\tilde{h_{0}}$ conjugate to the order parameter.
From the equation $\tilde{{\mathcal{M}}}_{2}=0$ we obtain the equation for the GL spinodal
curve
\[
\tilde{a}(\beta)=-2\frac{{\mathfrak{M}}_{2}^{(0)}}{{\mathfrak{M}}_{3}^{(0)}}+
\frac{{\mathfrak{M}}_{3}^{(0)}}{{\mathfrak{M}}_{4}^{(0)}},
\]
or
\begin{equation}
\frac{2}{\pi}\int_{0}^{\infty}\frac{x^{2}\sin x dx}{x^{3}T^{*}+24\eta\sin x}
=2\frac{S_{2}(0)}{S_{3}(0)}-\frac{S_{3}(0)}{S_{4}(0)},
\label{a3.14}
\end{equation}
where $S_{n}(0)$ is the $n$th structure factor of the one-component hard-sphere system
at $k=0$.

The phase diagram of the RPM  is shown in Fig.~2.
The curve with the maximum is the
GL spinodal calculated using (\ref{a3.14}). The Percus-Yevick
approximation is used  for $S_{2}(0)$ (the expressions for $S_{2}(0)$, $S_{3}(0)$ and
$S_{4}(0)$ are given in Appendix B). The straight line calculated by (\ref{a3.7})
(the Gaussian approximation) corresponds to
the charge ordering phase transition.
The GL critical point is located at $T_{c}^{*}=0.0502$ and $\eta_{c}=0.022$. While the
value for $T_{c}^{*}$ is in good agreement with the recent data of computer
simulations \cite{orkoulas1,caillol} ($T_{c}^{*}\simeq 0.05$), the critical density
is underestimated ($\eta_{c}\simeq 0.04$).

We can also obtain from (\ref{a3.13}) (using (\ref{a2.7})) the expression for $\mu_{+,1}$
\begin{equation}
\mu_{+,1}=-\frac{1}{2\beta}\frac{\mathfrak{M}_{2}^{(0)}\tilde{a}(\beta)-
\frac{1}{3}\frac{(\mathfrak{M}_{3}^{(0)})^{3}}{(\mathfrak{M}_{4}^{(0)})^{2}}}
{\mathfrak{M}_{2}^{(0)}+\frac{1}{2}\mathfrak{M}_{3}^{(0)}\tilde{a}(\beta)
-\frac{1}{2}
\frac{(\mathfrak{M}_{3}^{(0)})^{2}}{\mathfrak{M}_{4}^{(0)}}}.
\label{a3.15}
\end{equation}
 If we set $\mathfrak{M}_{3}^{(0)}=0$ the expression  (\ref{a3.15}) reduces to (\ref{a3.1}).

It should be pointed out that the above described scheme of
integration in the vicinity of the GL critical point cannot be
used in the region close to the line defined by (\ref{a3.7})
(dotted line in Fig.~2): the variables $\gamma_{\bf{k}}$ appear to
be the essential variables in this region.

\section{Conclusions}

We use the recently developed approach in the study of the critical behaviour of
ionic fluids. For the RPM we obtain the functional of the GPF in terms of the two
fluctuating fields, namely, the field $h_{0}$ conjugate to the total density fluctuations
and  the field $\gamma_{\bf{k}}$ conjugate to the charge density fluctuations.
The phase diagram of the system is calculated in the MF approximation
(that corresponds to setting ${\mathfrak{M}}_{n}^{(i_{n})}=0$ for $n\geq 3$ in the expression
for the chemical potential)
as well as taking into account the higher powers of the  field $h_{0}$. In the both cases
the phase diagram demonstrates the GL and charge ordering
phase instabilities. In the latter case the obtained value for the GL critical
temperature correlates well with the MC simulation data.

Following the scenario proposed in \cite{oksana, patkozmel}
for a binary fluid we integrate over $\gamma_{\bf{k}}$ (non-essential variables) with
the Gaussian density measure. As a result, we obtain an explicit expression for the
grand thermodynamic potential of the RPM in the vicinity of the GL critical point as an
expansion in terms of the field conjugate to the order parameter.
The expression implies the classical critical behaviour of the RPM.

\section*{Appendix~A}

Let us consider a classical two-component system of interacting particles
consisting of $N_{a}$ particles of species $a$ and $N_{b}$ particles of
species $b$. The system is in  volume $V$  at  temperature $T$.

        Let us assume that an  interaction  in  the  system  has a
pairwise additive character. The interaction potential between a
$\gamma$ particle at $\vec r_{i}$  and a $\delta$ particle at $\vec
r_{j}$  may be expressed as a sum  of two terms:
\begin{displaymath}
U_{\gamma\delta}(r_{ij})=\psi_{\gamma\delta}(r_{ij}) +
\phi_{\gamma\delta}(r_{ij}),
\end{displaymath}
where
$\psi_{\gamma\delta}(r)$  is a potential of a short-range repulsion
that can be chosen as an interaction between two hard spheres
$\sigma_{\gamma\gamma}$ and  $\sigma_{\delta\delta}$.
$\phi_{\gamma\delta}(r)$ is an attractive part of the potential which
dominates at large distances.

Let us start with a grand partition function of a two-component continuous system
\begin{equation}
\Xi=\sum_{N_{a}=0}^{\infty}\sum_{N_{b}=0}^{\infty}\prod_{\gamma=a,b}
\frac{z_{\gamma}^{N_{\gamma}}}{N_{\gamma}!}
\int(d\Gamma)
\exp\left[-\frac{\beta}{2}\sum_{\gamma\delta}\sum_{ij}
U_{\gamma\delta}(r_{ij})\right], \label{dA.001}
\end{equation}
%%%%%%%%%%%%%%%%%%%%%%%%%%%%%%%%%%%%%%%%%%%%%%%%%%%%%%%%%%%%%%%%%%%%%%%%%
where
$(d\Gamma)=\prod_{\gamma}d\Gamma_{N_{\gamma}}$,
$d\Gamma_{N_{\gamma}}=d\vec r_{1}^{\gamma}d\vec r_{2}^{\gamma}\ldots d\vec
r_{N_{\gamma}}^{\gamma}$
is an element of the configurational space of the $\gamma$th species;
$z_{\gamma}$ is the fugacity of the $\gamma$th species:
$z_{\gamma}=\exp(\beta\mu_{\gamma}^{'})$,
$\mu_{\gamma}^{'}=\mu_{\gamma}+\beta^{-1}\ln[(2\pi
m_{\gamma}\beta^{-1})^{3/2}/h^3]$;
$\beta=\frac{1}{k_{B}T}$,
$k_{B}$ is the Boltzmann constant, $T$ is temperature; $m_{\gamma}$
is mass of the $\gamma$th species, $h$ is the Planck constant.
$\mu_{\gamma}^{'}$ is determined from
\begin{displaymath}
\frac{\partial\ln\Xi}{\partial\beta\mu_{\gamma}^{'}} = \langle
 N_{\gamma}\rangle,
\end{displaymath}
where $\langle N_{\gamma}\rangle$ is the average number of
the $\gamma$th species.

     Further consideration of the problem is done in  the  extended
phase space: in the phase space of the Cartesian  coordinates  of  the
particles and in the CV phase space. An interaction connected with the
repulsion (potential $\psi_{\gamma\delta}(r)$) is  considered
in  the  space  of the Cartesian coordinates of the particles. We call
this two-component hard-spheres system a  reference  system  (RS).
The thermodynamic and structural properties of the RS are assumed to
be known. The interaction connected with an attraction (potential
$\phi_{\gamma\delta}(r)$ ) is considered in the CV space.
The phase space overflow is cancelled by introduction of a Jacobian of the
transition to CV. The contribution of the short-range forces to the
long-range interaction screening is  ensured by averaging this Jacobian over
the RS.

Let us introduce the grand partition function of the RS
\begin{equation}
\Xi_{0}=\sum_{N_{a}=0}^{\infty}\sum_{N_{b}=0}^{\infty}
\prod_{\gamma=a,b}
\frac{\exp{(\beta\mu_{0}^{\gamma}N_{\gamma})}}{N_{\gamma}!}
\int(d\Gamma)
\exp\left[-\frac{\beta}{2}\sum_{\gamma\delta}\sum_{ij}
\psi_{\gamma\delta}(r_{ij})\right], \label{dA.02}
\end{equation}
where $\mu_{0}^{\gamma}$ is the chemical potential of the $\gamma$th
species in the RS.

    Then the grand partition function (\ref{dA.001}) can be written
as \cite{patyuk3,patyuk4}:
\[
\Xi=\Xi_{0}\Xi_{1},
\]
%%%%%
where $\Xi_{0}$ is given in (\ref{dA.02}).
The part of the grand partition function which is  defined  in
the CV phase space has the form of the functional integral:
\begin{equation}
\Xi_{1}=\int(d\rho)exp[\beta\sum_{\gamma}\mu_{1}^{\gamma}\rho_{0,\gamma}-
\frac{\beta}{2V}\sum_{\gamma\delta}\sum_{\vec
k}\tilde\phi_{\gamma\delta}(k) \rho_{\bf{k},\gamma}\rho_{-\bf{k},
\delta}]J(\rho_{a},\rho_{b}). \label{dA.1}
\end{equation}
Here,

1) $\mu_{1}^{\gamma}$ is a part of the chemical potential of the
$\gamma$-th species
\begin{displaymath}
\mu_{1}^{\gamma}= \mu_{\gamma}-\mu_{0}^{\gamma} +
\frac{\beta}{2V}\sum_{\bf{k}}\tilde \phi_{\gamma\gamma}(k)
\end{displaymath}
and is determined from the equation
\begin{displaymath}
\frac{\partial\ln\Xi_{1}}{\partial\beta\mu_{1}^{\gamma}} = \langle
 N_{\gamma}\rangle,
\end{displaymath}
$\mu_{\gamma}$ is the full chemical potential of the $\gamma$-th
species;

2)$\rho_{\bf{k},\gamma}=\rho_{\bf{k},\gamma}^{c}-i\rho_{\vec
k,\gamma}^{s}$ is the collective variable of the $\gamma$-th species,
the indices $c$ and $s$ denote the real part and the coefficient at
the imaginary part of $\rho_{\bf{k},\gamma}$; $\rho_{\vec
k,\gamma}^{c}$ and $\rho_{\bf{k},\gamma}^{s}$  describe the value of
$\bf{k}$-th fluctuation mode of the number of $\gamma$-th  species
particles. Each of $\rho_{\bf{k},\gamma}^{c}$ and $\rho_{\vec
k,\gamma}^{s}$ takes all the real values from $-\infty$ to $+\infty$.
$(d\rho)$ is a volume element of the CV phase space:
\begin{displaymath}
(d\rho)=\prod_{\gamma}d\rho_{0,\gamma}{\prod_{\bf{k}\not=0}}'
d\rho_{\bf{k},\gamma}^{c}d\rho_{\bf{k},\gamma}^{s}.
\end{displaymath}
The prime means that the product over $\bf{k}$ is performed  in
the upper semispace;

3) $J(\rho_{a},\rho_{b})$ is the transition Jacobian to the
CV averaged on the RS \cite{patyuk3,patyuk4}:
\begin{eqnarray}
J(\rho_{a},\rho_{b}) & = & \int(d\nu)\prod_{\gamma=a}^{b}\exp\left
[i2\pi\sum_{\bf{k}} \nu_{\bf{k},\gamma}\rho_{\bf{k},\gamma}\right]
\exp\left [\sum_{n\geq 1}
\frac{(-i2\pi)^{n}}{n!}\times\right.\nonumber\\ &  & \left.
\sum_{\gamma_{1}\ldots\gamma_{n}} \sum_{{\bf{k}}_{1}\ldots{\bf{k}}_{n}}
M_{\gamma_{1}\ldots\gamma_{n}}({\bf{k}}_{1},\ldots,{\bf{k}}_{n}) \nu_{{\bf{k}}_{1},
\gamma_{1}}\ldots\nu_{{\bf{k}}_{n},\gamma_{n}}\right],
\label{dA.3}
\end{eqnarray}
where  variable $\nu_{\bf{k},\gamma}$ is
conjugated to  CV $\rho_{\bf{k},\gamma}$.
$M_{\gamma_{1}\ldots\gamma_{n}}({\bf{k}}_{1}, \ldots,{\bf{k}}_{n})$ is the
$n$-th cumulant connected  with
$S_{\gamma_{1}\ldots\gamma_{n}}(k_{1},\ldots,k_{n})$, the $n$-particle
partial structure factor of the RS, by means of the relation
\begin{displaymath}
M_{\gamma_{1}\ldots\gamma_{n}}({\bf{k}}_{1},\ldots,{\bf{k}}_{n})=
\sqrt[n]{N_{\gamma_{1}}\ldots
N_{\gamma_{n}}}S_{\gamma_{1}\ldots\gamma_{n}}
(k_{1},\ldots,k_{n})\delta_{{\bf{k}}_{1}+\cdots+{\bf{k}}_{n}},
\end{displaymath}
where $\delta_{{\bf{k}}_{1}+\cdots+{\bf{k}}_{n}}$ is the Kronecker symbol.

In general, the dependence of
$M_{\gamma_{1}\ldots\gamma_{n}}({\bf{k}}_{1}, \ldots,{\bf{k}}_{n})$ on
wave vectors ${\bf{k}}_{1},\ldots,{\bf{k}}_{n}$ is complicated. Hereafter we
shall replace $M_{\gamma_{1}\ldots\gamma_{n}}({\bf{k}}_{1}, \ldots,{\bf{k}}_{n})$
by their values in long-wave length limit
$M_{\gamma_{1}\ldots\gamma_{n}}(0,\ldots,0)$.

4)  $\tilde \phi_{\gamma\delta}(k)$ is the Fourier transform of
attractive potential $\phi_{\gamma\delta}(r)$. The
behaviour  of $\phi_{\gamma\delta}(r)$ in the region of the core
$r<\sigma_{\gamma\delta}$ must be determined from the conditions of
optimal separation of the interaction.

We pass in (\ref{dA.1}) to CV $\rho_{\bf{k}}$ and $c_{\bf{k}}$
(according to $\omega_{\bf{k}}$ and $\gamma_{\bf{k}}$) by means of the
orthogonal linear transformation:
\begin{displaymath}
\rho_{\bf{k}} =  \frac{\sqrt{2}}{2}(\rho_{{\bf{k}},a}+\rho_{{\bf{k}},b}),
\qquad
c_{\bf{k}} =  \frac{\sqrt{2}}{2}(\rho_{{\bf{k}},a}-\rho_{{\bf{k}},b}),
\end{displaymath}
\begin{displaymath}
\omega_{\bf{k}} =  \frac{\sqrt{2}}{2}(\nu_{{\bf{k}},a}+\nu_{{\bf{k}},b}),
\qquad
\gamma_{\bf{k}} =  \frac{\sqrt{2}}{2}(\nu_{{\bf{k}},a}-\nu_{{\bf{k}},b}).
\end{displaymath}
As a result, we obtain for $\Xi$ \cite{oksana,patkozmel}
\begin{eqnarray}
\Xi & = & \Xi_{0}\int
(d\rho)\,(dc)\exp\Big[\beta\mu_{1}^{+}\rho_{0}+\beta\mu_{1}^{-}c_{0}-
\frac{\beta}{2V}\sum_{\bf{k}}[\tilde V(k)\rho_{\bf{k}}\rho_{-\bf{k}}+
\nonumber \\
&&\tilde W(k)c_{\bf{k}}c_{-\bf{k}}+\tilde U(k)\rho_{\bf{k}}c_{-\bf{k}}]
\Big]J(\rho,c). \label{dA.4}
\end{eqnarray}
Here the following notations are introduced:

$\rho_{\bf{k}}$ and $c_{\bf{k}}$ are the CV connected with total
density fluctuation modes and relative density (or concentration)
fluctuation modes in the binary system.
$(d\rho)$ and $(dc)$ are volume elements of the CV phase space:
\[
(d\rho)=d\rho_{0}{\prod_{\bf{k}\neq 0}}^{'}d\rho_{\bf{k}}^{c}d\rho_{\bf{k}}^{s},
\qquad
(dc)=dc_{0}{\prod_{\bf{k}\neq 0}}^{'}dc_{\bf{k}}^{c}dc_{\bf{k}}^{s}.
\]
Functions $\mu_{1}^{+}$ and $\mu_{1}^{-}$ have the form:
\begin{equation}
\mu_{1}^{+}=\frac{\sqrt{2}}{2}(\mu_{1,a}+\mu_{1,b}),\qquad
\mu_{1}^{-}=\frac{\sqrt{2}}{2}(\mu_{1,a}-\mu_{1,b})
\label{dA.5}
\end{equation}
\begin{eqnarray}
\tilde V(k)&=&(\tilde \phi_{aa}(k)+\tilde \phi_{bb}(k)+2\tilde
\phi_{ab}(k))/2 \nonumber \\
\tilde W(k)&=&(\tilde \phi_{aa}(k)+\tilde \phi_{bb}(k)-2\tilde
\phi_{ab}(k))/2 \nonumber \\
\tilde U(k)&=&(\tilde \phi_{aa}(k)-\tilde \phi_{bb}(k))/2,
\label{dA.6}
\end{eqnarray}
\begin{equation}
J(\rho,c)=\int (d\omega)\,(d\gamma)\exp\Big[i2\pi\sum_{\bf{k}}
(\omega_{k}\rho_{k}+\gamma_{k}c_{k})\Big]J(\omega,\gamma),\label{dA.7}
\end{equation}
\begin{eqnarray}
J(\omega,\gamma)&=&\exp\Big[\sum_{n\geq 1}\sum_{i_{n}\geq
0}\frac{(-i2\pi)^{n}}{n!}\sum_{{\bf{k}}_{1}\ldots{\bf{k}}_{n}}
M_{n}^{(i_{n})}(0,\ldots,0)\times\nonumber \\
&  &\gamma_{{\bf{k}}_{1}}\ldots\gamma_{{\bf{k}}_{i_{n}}}
\omega_{{\bf{k}}_{i_{n+1}}}\ldots\omega_{{\bf{k}}_{n}}\Big].
\label{dA.8}
\end{eqnarray}
Index $i_{n}$ is used to indicate the number of variables
$\gamma_{\bf{k}}$ in the cumulant expansion (\ref{dA.8}). Cumulants
$M_{n}^{(i_{n})}$ are expressed as  linear combinations  of the partial
cumulants $M_{\gamma_{1}\ldots\gamma_{n}}$ (see (\ref{dA.3})) and
are presented for $\gamma_{1},\ldots,\gamma_{n}=a,b$  and $n\leq 4$ in
\cite{patyuk4} (see Appendix~B in \cite{patyuk4}).

\section*{Appendix~B}
\begin{eqnarray*}
S_{2}(0)&=&\frac{(1-\eta)^{4}}{(1+2\eta)^{2}},
\nonumber \\
S_{3}(0)&=&\frac{(1-\eta)^{7}(1-7\eta-6\eta^{2})}{(1+2\eta)^{5}},
\nonumber \\
S_{4}(0)&=&\frac{(1-\eta)^{10}(1-30\eta+81\eta^{2}+140\eta^{3}+60\eta^{4})}{(1+2\eta)^{8}}.
\end{eqnarray*}

\section*{Acknowledgement}
The author thanks M.F. Holovko for helpful discussions.

%%%%%%%%%%%%%%%%%%%%%%%%%%%%%%%%%%%%%%%%%%%%%%%%%555
\newpage

\begin{center}
FIGURE CAPTIONS
\end{center}

\begin{figure}
\centering
\includegraphics[height=6cm]{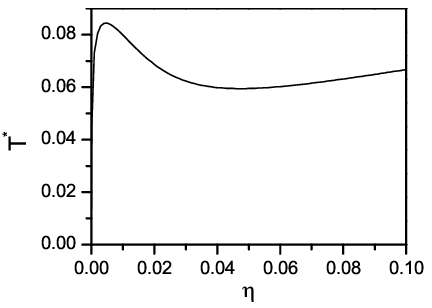}
\caption{The phase diagram of the RPM calculated from (\ref{a3.3})
(see the text for explanation).}
\label{fig1}
\end{figure}

FIG.~\ref{fig1}. The phase diagram of the RPM calculated from (\ref{a3.3})
(see the text for explanation).\\

\begin{figure}
\centering
\includegraphics[height=6cm]{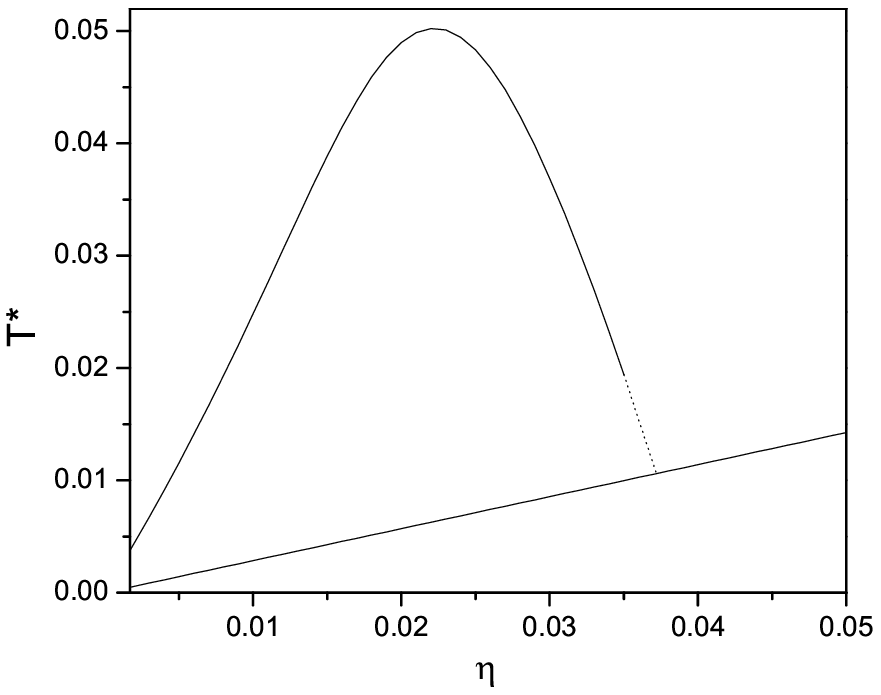}
\caption{The phase diagram of the RPM calculated from (\ref{a3.12})
(see the text for explanation).}
\label{fig2}
\end{figure}

FIG.~\ref{fig2}. The phase diagram of the RPM calculated from (\ref{a3.12})
(see the text for explanation).

\newpage

\clearpage


\begin{thebibliography}{99}
\bibitem{levelt1} M.H. Levelt Sengers and J.A. Given, Mol. Phys., {\bf 80}, 899, (1993).
%
\bibitem{pitzer} K.S. Pitzer, J. Phys.Chem, {\bf 99}, 13070, (1995).
\bibitem{fisher1} M.E. Fisher, J. Stat. Phys., {\bf 75}, 1, (1994).
%
\bibitem{fisher2} M.E. Fisher, J. Phys.: Condens. Matter, {\bf 8}, 9103, (1996).
%
\bibitem{stell1} G. Stell, J. Stat. Phys., {\bf 78}, 197, (1995).
%
\bibitem{stell2} G. Stell, J. Phys.: Condens. Matter, {\bf 8}, 9329, (1996).
%
\bibitem{zhou} Y. Zhou, S. Yeh and G. Stell, J.Chem.Phys., {\bf 102}, 5785, (1995).
%
\bibitem{fisher3} M.E. Fisher and B.P. Lee, Phys. Rev. Lett., {\bf 77}, 3361, (1996).
%
\bibitem{ciach1} A. Ciach and G. Stell, J Mol. Liq., {\bf 87}, 253, (2000).
%
\bibitem{ciachstell} A. Ciach and G. Stell, Physica A, {\bf 306}, 220, (2002).
%
\bibitem{caillol1} J.M. Caillol, D. Levesque and J.J. Weis,
Phys. Rev. Lett., {\bf 77}, 4039, (1996).
%
\bibitem{orkoulas1} G. Orkoulas and A.Z. Panagiotopoulos, J. Chem. Phys.,
{\bf 110}, 1581, (1999).
%
\bibitem{caillol}  Caillol J.-M., Levesque D., Weis J.-J.
Critical behavior of the
restricted primitive model revisited, Preprint cond-mat/0201301 v1.
%
\bibitem{luijten} E. Luijten, M.E. Fisher and A.Z. Panagiotopoulos, Phys. Rev. Lett.,
{\bf 88}, 185701-1, (2002).
%
\bibitem{panagiotopoulos1} A.Z. Panagiotopoulos, J. Chem. Phys.,
{\bf 116}, 3007, (2002).
%
\bibitem{oksana} O.V. Patsahan, Physica A, {\bf 272}, 358, (1999).
%
\bibitem{patkozmel} O.V. Patsahan ,M.P. Kozlovskii and  R.S. Melnyk,
J.Phys.:Condens. Matter, {\bf 12}, 1595, (2000).
%
\bibitem{zubar} D.N. Zubarev, DAN USSR, {\bf 95}, 757, (1964) (in Russian).
%
\bibitem{yukhol} I.R. Yukhnovskii and M.F. Holovko,  {\it Statistical
Theory of Classical Equilibrium Systems} (Naukova Dumka, Kiev, 1980).
(in Russian).
%
\bibitem{yuk} I.R. Yukhnovskii,{\it Phase Transitions of the Second Order:
Collective Variables Method} (World Scientific, Singapore, 1987).
%
\bibitem{yuk2}  I.R. Yukhnovskii, Proceedings of the Steclov Institute of
Mathematics, {\bf 2}, 223, (1992).
%
\bibitem{patyuk3} O.V. Patsagan and I.R. Yukhnovskii,
Teor. Mat. Fiz. {\bf 83}, 72, (1990) [Sov. Theoret. Math. Phys., {\bf 83}, 387, (1990)].
%
\bibitem{patyuk4} I.R. Yukhnovskii and O.V. Patsahan, J. Stat. Phys.,
{\bf 81}, 647, (1995).
%

\bibitem{patkozmel1} O.V. Patsahan, R.S. Melnyk and M.P. Kozlovskii,
Condens. Matter Phys., {\bf 4}, 235, (2001).
%
\bibitem{pat_pat} O.V. Patsahan and T.M. Patsahan, J. Stat.
Phys., {\bf 105}, 287, (2001).
%
\bibitem{stell3} Y. Zhou, S. Yeh and G. Stell, J. Chem. Phys., {\bf 102}, 5785, (1995).
%
\bibitem{wcha}
J.D. Weeks,  D. Chandler and H.C. Andersen, J. Chem. Phys., {\bf 54}, 5237, (1971).
%
\bibitem{cha}D. Chandler and H.C. Andersen, J. Chem. Phys., {\bf 54}, 26, (1971).
%
\bibitem{stell} G. Stell, in {\it The Equilibrium Theory of Classical Fluids}
edited by H. Frisch and J. Lebowitz (Benjamin, New York, 1964).
%
\end{thebibliography}
\end{document}